# A survey on Human Computer Interaction Mechanism Using Finger Tracking

Ms.Kinjal N. Shah[1], Mr.Kirit R. Rathod[2], Mr.Shardul J. Agravat[3]

[1](Computer Engineering Dept., P G Student, C U Shah College of Engineering and Technology, India)
[2](Computer Engineering Dept, Asst. Professor, C U Shah College of Engineering and Technology, India)
[3](Computer Engineering Dept, Asst. Professor, C U Shah College of Engineering and Technology, India)

**ABSTRACT:** Human Computer Interaction (HCI) is a field in which developer makes a user friendly system. User can interact with a computer system without using any conventional peripheral devices. Marker is used to recognize hand movement accurately & successfully. Researchers establish the mechanism to interact with computer system using computer vision. The interaction is better than normal static keyboard and mouse. This paper represents most of innovative mechanisms of the finger tracking used to interact with a computer system using computer vision.

## I. INTRODUCTION

Nowadays Computer indeed becomes a need to live life. Computer is a fixed set of peripherals to interact with the program and solve the human problems easily.

Human Computer Interaction (HCI) involves the planning and design of the interaction between users and computers. In these days, smaller devices are used to improve technology. The most important advantages of computer vision is its freedom. The user can interact with the computer without wires and manipulating intermediary devices. Recently, User-Interfaces are used to capture the motion of our hands. The researchers developed techniques to track the movements of hand/fingers through the web cam to establish an interaction mechanism between user and computer.

The technique of tracking the movement of fingers in front of a web cam is called finger tracking. We use a colored substance, motion detection, camera to control the mouse movement and implement the finger tracking.

For the proper gesture recognition various processes such as Segmentation, Background subtraction, Noise removal and Thresholding etc can be performed on each frame of video.

## II. COMPARISON OF VARIOUS APPROACHES USED FOR HCI

There are various approaches used for developing HCI through finger tracking using computer vision. These are as follows:

2.1 HCI without Using Interface:

There is no material used to interact the computer.

2.1.1 Plain Finger Tracking with Single Web Cam:

This approach uses the bare hand to track the movement of finger in front of the camera based on the skin color. This approach is based on low cost, but they need special lighting effect and condition related to background. When the lighting effect and background change, the result may change. The more accuracy is based on constant background and lighting condition.

2.2 HCI Using Interface:

Some materials like data gloves, markers etc used to interact the computers.

2.2.1 Pasting Marker on Finger:

We paste the color marker on the finger and then track the movement of the finger. In these, we have to find particular color of the marker from the frame. For finding the color, we have to perform some operation like Thresolding. This approach is better than plain finger tracking because it takes less time. If the background color is same as the marker color, then we can't detect the finger movement. So, we must take the static background to detect the finger movement easily and fast.

### 2.2.2 Using Gloves with Markers and a Simple Web Cam

In this, Simple cloth Gloves with specific color marker pasted on its finger. In this approach, markers are used to uniquely distinguish finger based on color of a marker. Static background always increases the accuracy. For each frame, we pre-defined the marker color. So, it reduces the processing time and gives the accurate result. This approach is used in cursor movement, virtual mouse etc.

### 2.2.3 Using Gloves with Retro Reflective Markers and an Infrared Web Cam:

Gloves are used with Retro reflective markers and Infrared Web cam. Infrared web cam is used to avoid lighting effect. This approach uses different lighting conditions. Infrared web cam can easily identify the Retro reflective markers. It increases the number of frames could be processed within a second. This approach is used in virtual reality.

### 2.2.4 Using Special Hardware:

This approach uses special hardware like charge coupled Digital camera, projector, Gloves etc. to track the movement of finger. This approach is used in 3D virtual environment. This approach uses the specialized hardware. So, it is costly and less used for design application.

## III. APPLICATION

Finger tracking through computer vision is mainly design new application which is free from the conventional interaction device.

### 3.1 Robot control

Robot control application is based on gesture recognition and counts the number of finger shown in front of camera [1].Robot follows the command which we give through the finger[1]. Robot will move forward, backward ,left or right according to the command which will given by finger.

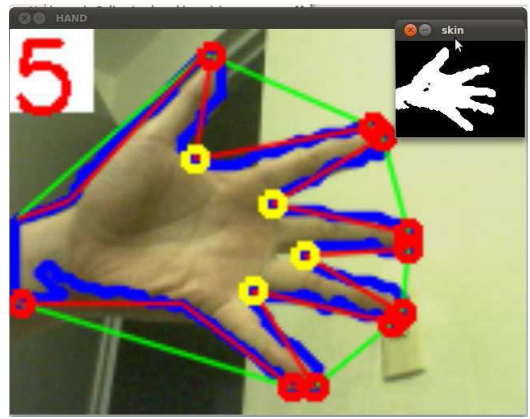

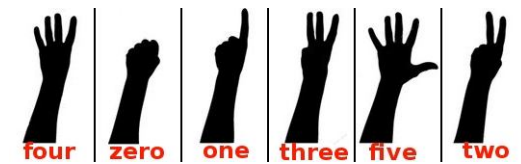

Figure: Robot control

### 3.2 Finger Painting

In finger painting, we can track the particular location of the finger and using that location we can create a painting[4]. We can generate a digital drawing according to the movements of fingers happening in front of a webcam. This is useful to the painters to create their painting directly in the digital form.

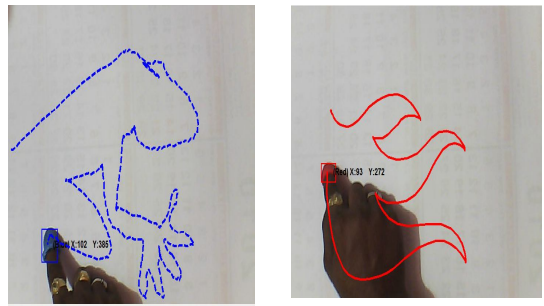

Figure: Finger Painting

### 3.3 Piano application

In this application, the user track the finger's position and pointing out the switches[9]. With the help of the finger pointing, we can create the sound of the piano.

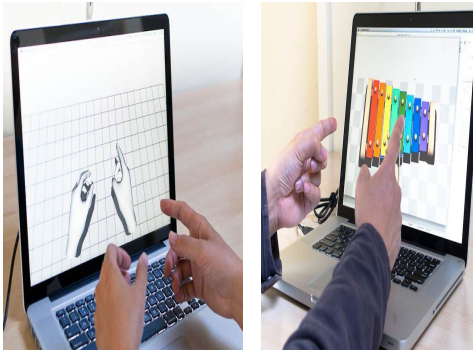

Figure: Piano application using finger tracking

### 3.4 Free Air Finger Tracking

Free air finger tracking basically contains the zoom in, zoom out and clicking event. Finger tracking can drag the pictures and also changing the size of pictures[10]. We can click and open the data with the help of free air finger tracking.

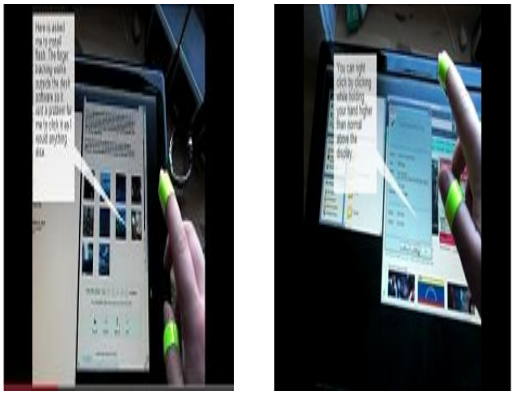

Figure: Free Air Finger Tracking

### 3.5 Remote Control

Remote control is an application to control a TV set. By moving the user's finger, a user could control the select channels [7]. We can also use this remote control application for CD player or any other remote devices.

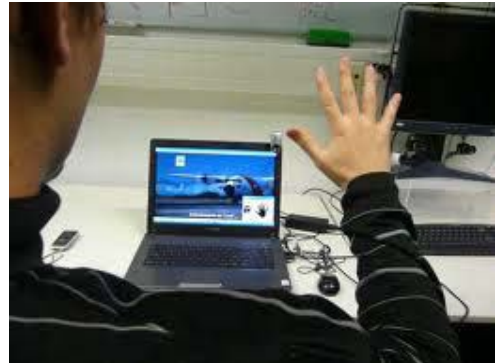

Figure: Remote Control

### 3.6 Interacting with a Virtual Environment

In the Virtual Environment, We interact with the virtual object [9]. The user feels, he/she is actually interacting with the real object, while interacting with the digital objects inside the computer system. Nowadays with the help of Virtual Environment, We create virtual games, driving simulations, animation movies etc.

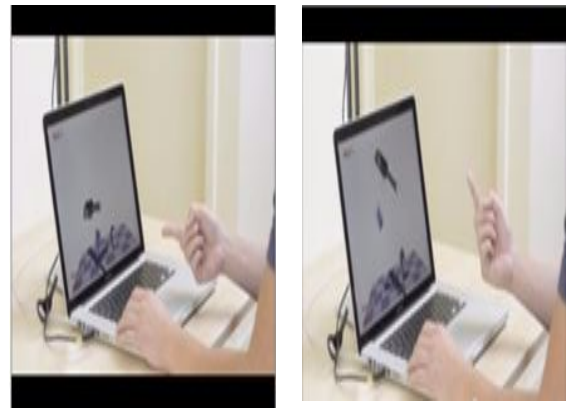

Figure: Virtual Reality based Game

### 3.7 Augmented Reality

Augmented Reality has been demonstrated widely in this real world. Augmented reality used with many different applications such as games, navigation and references. Augmented reality applications became increasingly interactive. Augmented Reality works on the successful demonstration of direct free-hand gestures.

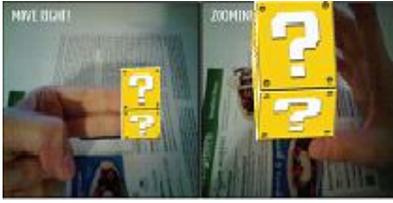

Figure: Augmented Reality based scalling

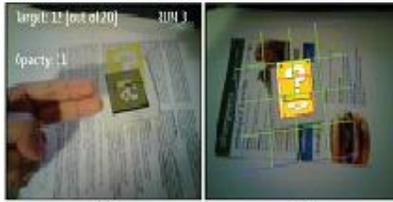

Figure: Augmented Reality in the 4 * 4 grid board we use to track the hand.

## IV. CONCLUSION

Human Computer Interaction (HCI) through computer vision is more convenient way to interact with the device. Numbers of application for the finger tracking are developed by researchers. They always try to use minimize the number of peripheral devices to interact with the computer. So, it is more useful for the physically disabled people. We expect that, the researcher will get extra ordinary application as an outcome of this effort.